\documentclass[aps,prl,twocolumn,groupedaddress]{revtex4}
\usepackage{epsfig}
\usepackage{graphics}

\begin{document}

\title{Photocontrol of Protein Conformation in a Langmuir Monolayer}

\author{Pietro Cicuta}
\author{Ian Hopkinson} 
\author{Peter G. Petrov}\email[Corresponding author. Present address: School of Physics, University of Exeter, Stocker Road, Exeter EX4 4QL, U.K.]{p.g.petrov@exeter.ac.uk}
\affiliation{Cavendish Laboratory, University of Cambridge, Madingley Road, Cambridge CB3 0HE, U.K.}

\begin{abstract} 
We report a method to control the conformation of a weak polyampholyte (the
protein $\beta$-casein) in Langmuir monolayers by light, even though the
protein is not photosensitive. Our approach is to couple the monolayer state to
a photochemical reaction excited in the liquid subphase. The conformational
transition of the protein molecule is triggered through its sensitivity to a
subphase bulk field (pH in this study), changing in the course of the
photochemical process. Thus, reaction of photoaquation of the ferrocyanide ion,
which increases the subphase pH from 7.0 to about 8.3, produces a change in the
surface monolayer pressure, $\Delta\Pi$, between -0.5 and +1.5 ${\rm mN/m}$
(depending on the surface concentration), signalling a conformational switch.
The approach proposed here can be used to selectively target and influence
different interfacial properties by light, without embedding photosensitisers
in the matrix. 
\end{abstract}


\maketitle

\section{Introduction}

It is well known that the state of monolayers spread on a liquid support can be
changed by light. So far, the approach has been to prepare Langmuir monolayers
of specially synthesised photochromic substances able to isomerise under
illumination with light of a suitable wavelength. Light-induced {\em cis-trans}
isomerisation of chromophores based on azobenzene and stilbene has often been
used \cite{Blair80,Karthaus96,Seki98,Kim00,Sidorenko00} and large changes in
the monolayer pressure-area isotherms have been achieved \cite{Karthaus96}.
Another extensively exploited isomerisation reaction is based on the
photoconversion of spiropyran chromophores into merocyanine species
\cite{Polymeropoulos79,Gruler80,Vilanove83,Moebius84,Panaiotov91}. In this
paper we propose a different approach, where the properties of the bulk liquid
subphase are changed by a photochemical reaction. Using the sensitivity of the
monolayer molecules to changes in the subphase, an indirect photoresponse is
triggered in the monolayer. Thus, it is not necessary to chemically modify the
monolayer in order to achieve a photocontrolled response. Following this
approach, we demonstrate here how the rich conformational behaviour of a
protein in a Langmuir monolayer can be controlled and directed non-invasively
by light. 

\section{Photochemistry}

We used the photoaquation \cite{Moggi66} of hexacyanoferrate (II) ion, ${\rm
Fe(CN)_6^{4-}}$, to change the pH of the liquid subphase beneath a
$\beta$-casein monolayer. Under illumination in aqueous solution, one of the
cyanide ions is released from the co-ordination sphere of ${\rm Fe^{2+}}$ and
substituted by a water molecule to give aquapentacyanoferrate (II), ${\rm
Fe(CN)_5H_2O^{3-}}$. The released ${\rm CN^-}$ is protonated to the weak
hydrocyanic acid, ${\rm HCN}$, which causes a pH increase in the solution
\cite{Mori94}. These processes are reversible when the illumination is stopped.
In this way, ferrocyanide photoaquation may be used to photochemically control
pH. Recently, a similar concept was used in order to tune the effective
spontaneous curvature of giant phospholipid vesicles \cite{Petrov99}. 

We measured photometrically the light-induced pH change of 1 mM potassium
hexacianoferrate (II) solution. Upon irradiation,the pH increases monotonically
from 7.0 to about 8.3. After turning off the light, the pH decreases to a constant
value, slightly higher (by about 0.1) than the initial ``dark'' pH. It has been shown \cite{Mori94} that a number of additional chemical processes can be expected (e.g., dimerisation of ${\rm Fe(CN)_5H_2O^{3-}}$, acid-base equilibrium of the dimer, etc.) which have resulted in a non-monotonic pH change with time. These, however, take place at longer illumination times and slightly higher concentrations. The monotonic pH increase during illumination, observed by us, is a strong indication that the ferrocyanide aquation and cyanide protonation are the primary photochemical processes under our experimental conditions.  

\section{Protein Monolayer Characterisation}

The protein $\beta$-casein is a weak polyampholyte with an amino acid sequence
with basic and acidic groups, whose charge can be adjusted by pH. Its
isoelectric point ${\rm pI} \approx 5$. The N-terminal region is more
hydrophilic than the rest of the chain, because many of the groups bear charge
\cite{Schwenke98}. When the protein is spread as a Langmuir monolayer, this
hydrophilic part would be expected to extend into the subphase to form a
``tail'' \cite{Graham79,Dickinson88,Leermakers96}. 

We characterised the protein monolayer by recording the surface pressure -
surface concentration isotherms, $\Pi(\Gamma)$, using a Langmuir trough. Figure
1a
\begin{figure}
\epsfig{file=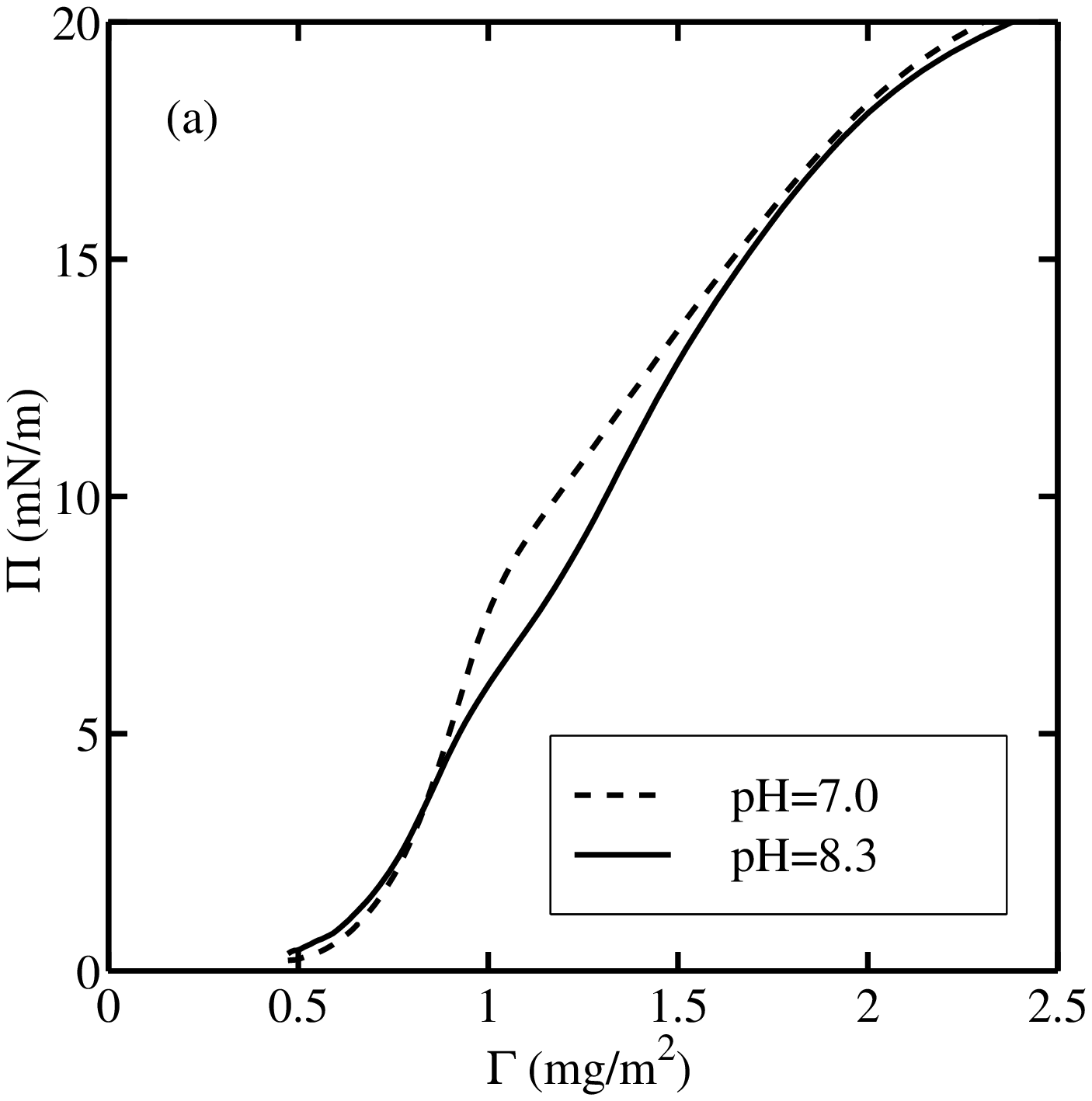,width=8cm}\\
\epsfig{file=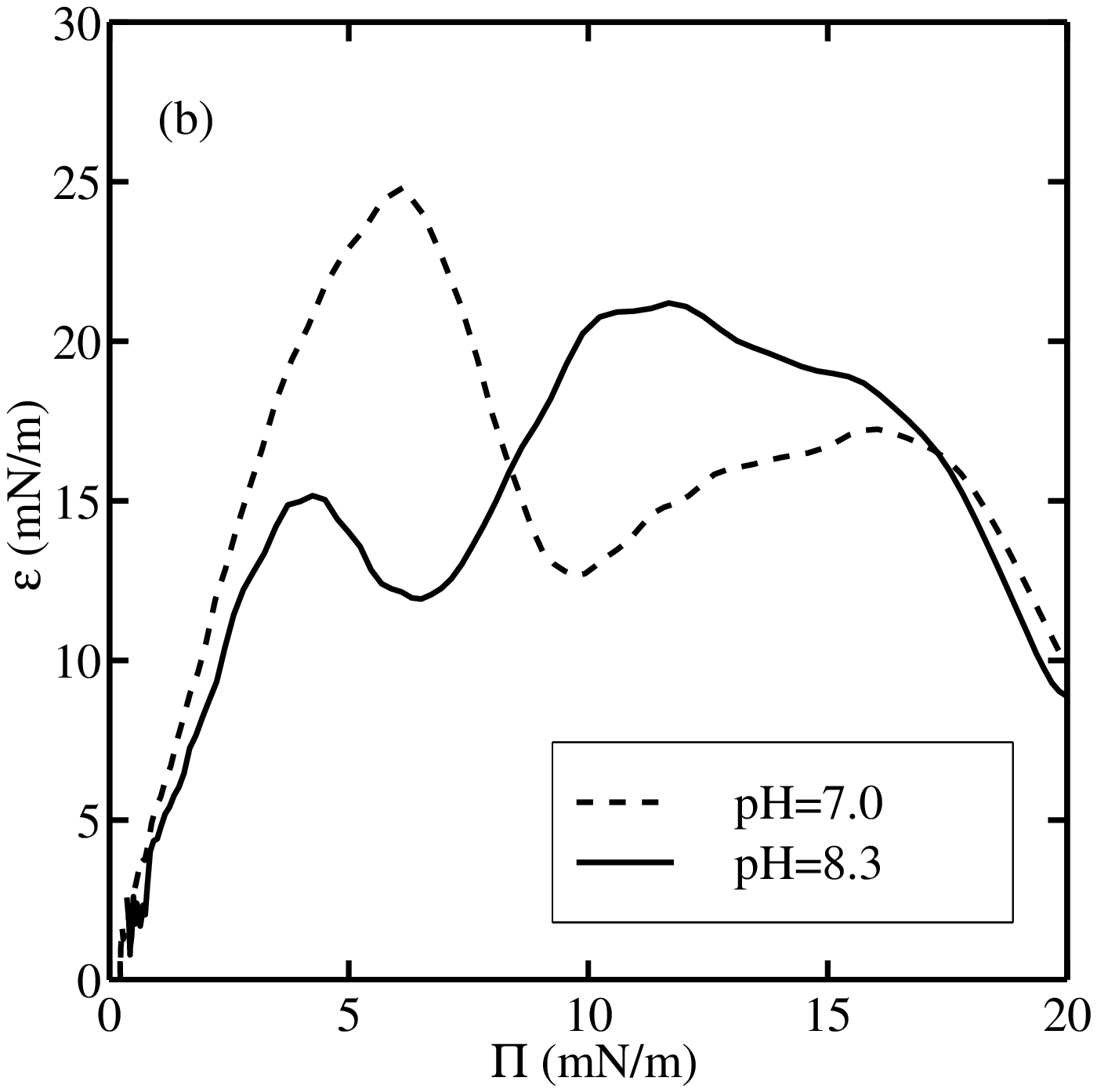,width=8cm}
\caption{\label{fig1}  (a) Surface pressure-surface concentration isotherms of
$\beta$-casein monolayer at pH = 7 (dashed line) and pH = 8.3 (solid line). The
subphase pH was adjusted by 100 mM phosphate buffer. (b) Static dilational
modulus, $\epsilon$, as a function of the surface pressure, $\Pi$, as
calculated from Figure 1(a).  }
\end{figure}
shows the results for two different subphase pH values, 7.0 and 8.3,
corresponding to the expected change in the pH due to the photochemical
reaction. Figure 1b represents the static dilational modulus, $\epsilon =
\Gamma(d\Pi/d\Gamma)$, as a function of $\Pi$, extracted from Figure 1a. At low
surface coverage, $\Gamma < 0.8~{\rm mg/m^2}$, the protein molecules are
assumed to be adsorbed entirely on the liquid surface in train conformation
with no tails or loops protruding into the liquid phase. Here, the surface
pressure at ${\rm pH} = 8.3$ is slightly, but distinguishably higher than the
surface pressure at neutral pH (see Figure 1a). This reflects the fact that the
polyampholyte molecule is more swollen at higher pH due to increased charge
density. Conformational changes, such as tail protrusion in the liquid and/or
formation of loops, are expected when the monolayer is compressed, depending on
the subphase pH and ionic strength \cite{Graham79,Dickinson88,Cicuta01}. It has
been proposed that a maximum in the $\epsilon~vs.~\Pi$ dependence (Figure 1b)
could be considered as an onset of a conformational transition and the
following minimum is the end of that transition
\cite{Graham79,Cicuta01,Mellema98}. On this basis, the first maximum (at $\Pi
\approx 6$ mN/m for ${\rm pH} = 7$ and $\Pi \approx 4$ mN/m for ${\rm pH} =
8.3$) appears to reflect the onset of tail formation, due to the more
hydrophilic N-terminus of the $\beta$-casein molecule
\cite{Cicuta01,Mellema98}. The second maximum, at $\Pi \approx 16$ mN/m for
${\rm pH} = 7$ and $\Pi \approx 11$ mN/m for ${\rm pH} = 8.3$, has been tentatively 
assigned to the onset of loop formation \cite{Cicuta01}. Direct structural probe techniques to resolve the surface conformation of this protein, like neutron \cite{Dickinson93} or X-ray reflectivity, have been limited to high surface concentrations. Only recently, Harzallah et al. \cite{Harzallah98} were able to achieve enough sensitivity at relatively low surface coverage between 1.14 and 2.69 ${\rm mg/m^2}$ for adsorption layers of $\beta$-casein. At the lowest $\Gamma = 1.14~{\rm mg/m^2}$ and ${\rm pH}=7.1$, the molecules were accomodated on the liquid surface as trains (56 \%) and loops or the N-tail (44 \%) with area per molecule $\approx 3500~ {\rm\AA}^2$. In contrast, at higher concentrations ($\Gamma > 2.13 ~{\rm mg/m^2}$), the train fraction decreased to about 30 \%, and the rest 70 \% of the protein sequence was present as loops and long tails. This, together with a two-fold decrease in the area per molecule to $\approx 1870 ~{\rm\AA}^2$, is in unison with the simplified conformational picture conjectured on the basis of thermodinamic ($\Pi,~\epsilon$) measurements. A detailed discussion of the conformational behaviour in $\beta$-casein monolayers can be found in \cite{Cicuta01}.   

\section{Photocontrol of Protein Conformation}

The experimental set up to induce conformational transitions of the protein
monolayer is straightforward. We chose to work with small surface areas in
order to achieve more intense and homogeneous illumination  and used a small
circular vessel (Petri dish of 6.8 cm diameter) as the Langmuir trough. The
light source, a 100 W halogen lamp, was mounted about 15 cm from the liquid
surface to produce unattenuated homogeneous illumination. The power density,
measured using a calibrated photodiode after a 505-575 nm bandpass filter, was
18 ${\rm mW/cm^2}$. The protein monolayer was spread on a subphase containing 1
mM potassium hexacyanoferrate (II) and 100 mM NaCl. The surface pressure was
recorded by a Wilhelmy plate. For each experiment, a new monolayer of different
surface concentration was spread, which was equivalent to the standard Langmuir
trough procedure, where the surface concentration is altered by changing the
surface area \cite{Gau94}. This procedure also ensured a standard initial state
of the monolayer under ``dark'' conditions. After recording the surface
pressure, the photochemical reaction was initiated by turning the illumination
on and the surface pressure relaxation was followed.

We recorded light-induced jumps in the surface pressure spanning the entire
region of surface concentrations from 0 to about 2.2 ${\rm mg/m^2}$. Four
examples, for different concentration regions, are shown in Figure 2.
\begin{figure}
\epsfig{file=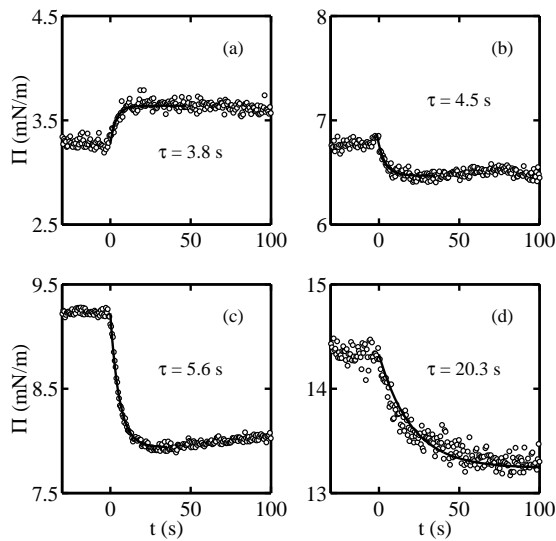,width=8cm}%
\caption{\label{fig2}  Four examples of light-induced surface pressure jumps for
different initial surface concentrations, signalling different types of
conformational switch. The surface concentration and the corresponding initial
surface pressure increase from (a) to (d). The relaxation times $\tau$ were
extracted by exponential fits (solid lines) to the experimental data.  }
\end{figure}
Generally, we observed fast changes in the surface pressure before it levelled
off to a plateau (see Figure 2). Often, especially at higher surface pressures,
the system exhibited a  more complicated pressure trend over a longer time scale
(data not shown) most probably due to stress redistribution in the viscoelastic
protein network. 

Two additional control experiments were performed. First, we checked the impact
of the heat produced by the light source in a system without the photochemical
compound. This produced a negligible change in the surface pressure. The second
experiment was performed in the presence of ferrocyanide, but this time in a pH
buffered subphase, ensuring that although the photochemical reaction occurred,
there was no change in pH. Again, no change in the surface pressure was
observed during illumination. This is an evidence that the primary
photochemical coupling is by the photo-induced pH change. Photoinduced changes
in the ionic strength were negligible, compared to the ionic strength set by
the inert monovalent electrolyte used throughout all the experiments (100 mM
NaCl).

Figure 3
\begin{figure}
\epsfig{file=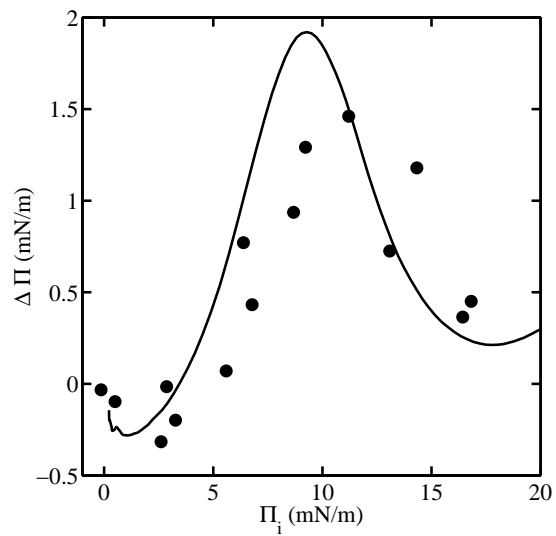,width=8cm}%
\caption{\label{fig3}  Dependence of photoinduced surface pressure jump, $\Delta \Pi =
\Pi_i - \Pi_f$, on the initial monolayer surface pressure in ``dark'', $\Pi_i$
(solid circles). The solid line represents the difference in the monolayer
surface pressures for two different values of the subphase pH, 7.0 and 8.3,
calculated from Figure 1a.  }
\end{figure}
 shows the dependence of the photoinduced surface pressure jump,
$\Delta \Pi$, on the initial surface pressure under ``dark'' conditions,
$\Pi_i$. The solid line shows the expected jump in the surface pressure due to
the change in the subphase pH, obtained from the pressure-concentration
isotherms in Figure 1. The agreement is fairly good, with the photochemical
system recovering all the important features, particularly the negative value
of $\Delta \Pi$ at very low initial pressures and the presence and approximate
position of the maximum. 

The protein behaviour can now be discussed on the basis of the proposed
molecular conformation as evidenced by the change in the static dilational
modulus due to pH changes (see Figure 1b and its discussion above). At low
surface concentrations (corresponding to low initial surface pressures),
irradiation leads to increase in the surface pressure (Figure 2a) and $\Delta
\Pi$ is negative. Here, the molecules are in all-train conformation. The
light-induced pH increase causes swelling of the molecules due to their
charging and this leads to increased surface pressure. With the increase of the
initial pressure, a cross-over to positive $\Delta \Pi$ is observed and the
photoinduced conformational switch has a different character. The pressure
relaxations in Figures 2b and 2c reflect transitions from train to tail
conformations (cf Figure 1b). That of Figure 2d can be assigned to a
light-induced formation of loops, i.e., transition from tail to tail-and-loop
conformation.

All data for the surface pressure relaxation are described well by a single
relaxation time, $\tau$, using the exponential decay function $\Pi(t)-\Pi_f=\Delta
\Pi ~{\rm exp}(-t/\tau)$  (where $\Pi_f$ is the plateau value of the surface
pressure). The exponential relaxation in this system can be understood as a
first-order kinetics assuming a rate of conformational reorientation
proportional to the instantaneous number of molecules to be converted
\cite{Balashev97}. Further, we found an increase of the relaxation time with
the increase of the initial pressure $\Pi_i$ (Figure 4).
\begin{figure}
\epsfig{file=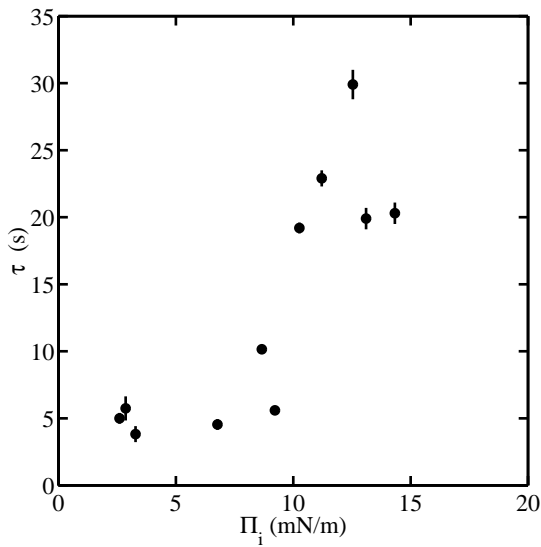,width=8cm}%
\caption{\label{fig4}  Dependence of the relaxation time, $\tau$, on the initial
monolayer surface pressure in ``dark'', $\Pi_i$.   }
\end{figure}
 This fact makes clear
that the slowest and therefore rate-determining step in the whole chain of
processes (photoinduced pH increase $\rightarrow$ charging of molecules
$\rightarrow$ conformational transition) is the conformational  rearrangement
in the protein monolayer. Obviously, the $\tau(\Pi_i)$ trend is due to steric
hindrance and entanglement of the protein molecules at higher surface
concentrations, increasing the energy barrier for conformational transition. By
way of contrast, a recent study of {\em cis-trans} photoisomerisation kinetics
in Langmuir monolayers of azobenzene-containing dendrons showed the opposite
trend: the isomerisation was facilitated at higher pressures, probably due to
collective behaviour \cite{Sidorenko00}.    

Finally, we would like to discuss the reversibility of the photoinduced
conformational transitions. As mentioned earlier, the ferrocyanide
photoaquation is reversible upon turning the illumination off. Our experimental
results show that full recovery of the surface pressure after ceasing the
illumination is possible at low surface concentrations, where the only process
induced is the  swelling/deswelling of the polyampholyte chain. The system
exhibits interesting behaviour in the intermediate region of the initial
surface pressures between 5 and 8 mN/m. Light-induced transitions here are
almost completely irreversible (data not shown) and the monolayer appears to be
trapped in a metastable state after the illumination. This corresponds,
according to Figure 1, to the only region of surface concentrations, where both
the pressure and the dilational modulus must increase if the monolayer were to
relax to its initial condition after the illumination has been terminated. At
higher surface concentrations (initial pressures above 8 mN/m), the
conformational transitions are, at best, only partially reversible, despite the
practically full recovery of the initial ``dark'' value of the subphase pH.
This situation is illustrated in Figure 5.
\begin{figure}
\epsfig{file=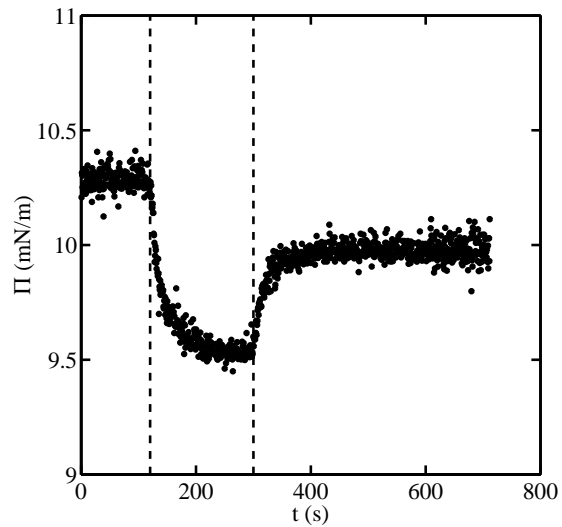,width=8cm}%
\caption{\label{fig5}  Reversibility of the photo-induced conformational switch. The
illumination has been turned on at $t = 120$ s and off at $t = 300$ s (see the
dashed lines).   }
\end{figure}
 Another factor affecting the
conformational reversibility is the exposure time. Longer illumination times
appear to make the light-induced conformational transition irreversible.  

\section{Summary and Conclusions}

In summary, we have demonstrated a novel method to control the molecular
conformation in Langmuir monolayers by coupling the monolayer state to a
photochemical reaction taking place in the bulk of the liquid subphase. We have
shown how different types of conformational switch (swelling of the
polyampholyte chain, train-to-tail and tail-to-tail-and-loop transitions) can
be induced by light by selecting the surface concentration. We consider the
present approach as a particular case within a more general scheme of coupling
between ongoing bulk chemistry and interfacial material properties and
topology, manifested in many biological and technologically important
processes. Identification of suitable couplings, especially to photochemical
and oscillating chemical reactions, would be of particular interest in this
respect.  

\begin{acknowledgments}
Useful discussions with H.-G. D{\"o}bereiner, E. M. Terentjev, M. Warner and S.
M. Clarke are greatly appreciated. This work has been financially supported by
EPSRC UK.   
\end{acknowledgments}

\end{document}